\documentclass[english,12pt]{article}
\usepackage{array}
\usepackage{graphicx}
\usepackage{amssymb}
\usepackage{amsmath}
\usepackage{multirow}
\usepackage{prettyref}
\usepackage{babel}
\usepackage{units}
\usepackage[latin1]{inputenc}
\usepackage{amsfonts}
\usepackage{amssymb}
\usepackage{float}
\usepackage{cancel}
\usepackage{dsfont}
\usepackage{cite}
\def\@fmsl@sh#1#2#3{\m@th\ooalign{$\hfil#1\mkern#2/\hfil$\crcr$#1#3$}}
 \def\eq#1\en{\begin{equation}#1\end{equation}}
\def\s[#1,#2]{[#1\stackrel{\star}{,}#2]}
\def\sx[#1,#2]{[#1\stackrel{\star_{x}}{,}#2]}

\newcommand{\Rlog}[1]{\ln\left(\cfrac{#1}{\mu^{2}}\right)}
\newcommand{\pd}{\partial}
\newcommand{\nc}{\newcommand}
\nc{\beq}{\begin{equation}}
\nc{\eeq}{\end{equation}}
\nc{\beqa}{\begin{eqnarray}}
\nc{\eeqa}{\end{eqnarray}}

\def\bc{\begin{center}}
\def\ec{\end{center}}

\def\gsim{\mathrel{\mathpalette\atversim>}}

\def\bc{\begin{center}}
\def\ec{\end{center}}

\def\gsim{\mathrel{\rlap{\lower4pt\hbox{\hskip1pt$\sim$}}

    \raise1pt\hbox{$>$}}}       

\def\gsim{\mathrel{\rlap{\lower4pt\hbox{\hskip1pt$\sim$}}
    \raise1pt\hbox{$>$}}}       

\begin{document}
\makeatletter
\def\fmslash{\@ifnextchar[{\fmsl@sh}{\fmsl@sh[0mu]}}
\def\fmsl@sh[#1]#2{%
  \mathchoice
    {\@fmsl@sh\displaystyle{#1}{#2}}%
    {\@fmsl@sh\textstyle{#1}{#2}}%
    {\@fmsl@sh\scriptstyle{#1}{#2}}%
    {\@fmsl@sh\scriptscriptstyle{#1}{#2}}}
\def\@fmsl@sh#1#2#3{\m@th\ooalign{$\hfil#1\mkern#2/\hfil$\crcr$#1#3$}}
\makeatother

\thispagestyle{empty}
\begin{titlepage}
\boldmath
\begin{center}
  \Large {\bf Gravity Induced Non-Local Effects in the Standard Model}
    \end{center}
\unboldmath
\vspace{0.2cm}
\begin{center}
{{\large S.~O.~Alexeyev$^{a,}$}\footnote{alexeyev@sai.msu.ru},
  {\large X.~Calmet$^{b,c,}$}\footnote{x.calmet@sussex.ac.uk} {\large and}
{\large B.~N.~Latosh$^{b,d,e,}$}\footnote{b.latosh@sussex.ac.uk}}
 \end{center}
\begin{center}
{\sl $^a$Sternberg Astronomical Institute, Lomonosov Moscow State University, Universitetsky Prospekt, 13, Moscow 119991, Russia}\\
{\sl $^b$Department of Physics and Astronomy, 
University of Sussex,   Falmer, Brighton, BN1 9QH, United Kingdom 
}\\
$^c${\sl PRISMA Cluster of Excellence and Mainz Institute for Theoretical Physics, Johannes Gutenberg University, 55099 Mainz, Germany }\\
{\sl $^d$Faculty of Natural and Engineering Science, Dubna State University, Universitetskaya Str. 19, 141980 Dubna, Moscow Region, Russia\\ 
$^e$Bogoliubov Laboratory of Theoretical Physics, Joint Institute for Nuclear Research, Joliot-Curie 6, 141980 Dubna, Moscow Region, Russia}
\end{center}
\vspace{5cm}
\begin{abstract}
\noindent
We show that the non-locality recently identified in quantum gravity using resummation techniques propagates to the matter sector of the theory. We describe these non-local effects  using effective field theory techniques. We derive the complete set of non-local effective operators at order $N G^2$ for theories involving scalar, spinor, and vector fields. We then use recent data from the Large Hadron Collider to set a bound on the scale of space-time non-locality and find $M_\star> 3 \times 10^{-11}$ GeV. 
\end{abstract}  
\end{titlepage}


Finding a quantum mechanical description of General Relativity, in other words, a quantum theory of gravity, remains one of the holy grails of modern theoretical physics. While it is not clear what this fundamental theory might be, we can use effective theory techniques to describe quantum gravity at energies below the Planck scale $M_P=1/\sqrt{G}$ where $G$ is Newton's constant. This approach is justified by the requirement that whatever the correct theory of quantum gravity might be, General Relativity must arise in its  low energy limit.

We do not have much information about physics at the Planck scale as experiments at this energy scale are difficult to imagine. We, nevertheless, have indications that a unification of General Relativity and Quantum Mechanics may lead to a more complicated structure of space-time at short distances in the form of a minimal length. Indeed, there are several thought experiments \cite{Mead:1964zz,Padmanabhan:1987au,Garay:1994en,Salecker:1957be,Calmet:2004mp,Calmet:2005mh,Calmet:2007vb} showing that, given our current understanding of Quantum Mechanics, General Relativity and causality, it is inconceivable to measure distances with a better precision than the Planck length $\l_P=\sqrt{\hbar G/c^3}$ where $\hbar$ is the reduced Planck constant and $c$ is the speed of light in vacuum. Such arguments imply a form of non-locality at short distances of the order of $l_P$. We will show that the scale of non-locality could actually be much larger that $l_P$ depending on the matter content in the theory.

 An important question is whether this non-locality could be found when combining quantum field theory with General Relativity as well. In \cite{Calmet:2015dpa}, it was shown that General Relativity coupled to a quantum field theory generically leads to non-local effects in scalar field theories.  In the current paper, we build on the results obtained in \cite{Calmet:2015dpa} and extend them to matter theories involving spinor and vector fields as well. We show that non-local effects are universal and affect all matter fields. We derive a complete set of non-local effective operators at order $N G^2$ where $N=N_s +3 N_f +12 N_V$ with $N_s$, $N_f$ and $N_V$ denoting respectively  the number of scalar, spinor, and vector fields in the theory. Then, using  recent data from the Large Hadron Collider, we set a limit on the scale of space-time non-locality.

\begin{figure}[htp]
\centering
\includegraphics[ width=5in]{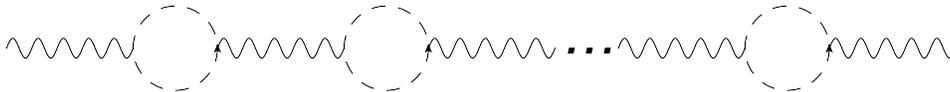}
\caption{Resummation of the graviton propagator.}\label{Fig1}
\end{figure}

Recently, several groups have studied perturbative linearized General Relativity coupled to matter fields. They found that perturbative unitarity can breakdown well below the reduced Planck mass \cite{Han:2004wt,Atkins:2010eq,Atkins:2010re,Atkins:2010yg}. The self-healing mechanism \cite{Aydemir:2012nz,Calmet:2013hia} demonstrates that unitarity can be recovered by resumming a series of graviton vacuum polarization diagrams in the large $N$ limit (Fig. (1)), see as well  \cite{Tomboulis:1977jk,Tomboulis:1980bs} for earlier works on large $N$ quantum gravity.
An interesting feature of this large $N$ resummation, while keeping $N G$ small, is that the obtained resummed graviton propagator 
\begin{eqnarray} \label{resprop}
i D^{\alpha \beta,\mu\nu}(q^2)=\frac{i \left (L^{\alpha \mu}L^{\beta \nu}+L^{\alpha \nu}L^{\beta \mu}-L^{\alpha \beta}L^{\mu \nu}\right)}{2q^2\left (1 - \frac{N G q^2}{120 \pi} \log \left (-\frac{q^2}{\mu^2} \right) \right)},
\end{eqnarray}
where $\mu$ is the renormalization scale incorporates some of the non-perturbative physics of quantum gravity. It has poles beyond the usual one at $q^2=0$.  Indeed, one finds \cite{Calmet:2014gya,Calmet:2015pea,Calmet:2016sba} that there is a pair of complex poles at 
\begin{eqnarray}
q^2&=& \frac{1}{G N} \frac{120 \pi}{ W\left (\frac{-120 \pi}{\mu^2 N G} \right)}
\end{eqnarray}
where $W$ is the Lambert function.  As explained in \cite{Calmet:2014gya}, these complex poles are a sign of strong interactions.   The mass and width of these objects can be calculated. It was suggested in \cite{Calmet:2014gya} that the complex poles could be interpreted as black hole precursors. These Planckian black holes are purely quantum object and their geometry is not expected to be described accurately by the standard solutions of classical Einstein's equations. In particular, they will not decay via Hawking radiation as they are non-thermal objects. While they do not radiate, they are very short-lived objects and will decay to a few particles.  Their widths are of the order of $(120 \pi/ G N)^{1/2}$. Because the complex poles are related by complex conjugation, one of them has an incorrect sign between its mass and its width and it corresponds to a particle propagating backwards in time. This complex pole thus leads to acausal effects which should become appreciable at energies near $(120 \pi/ G N)^{1/2}$. Using the in-in formalism \cite{Schwinger:1960qe,Keldysh:1964ud} it is possible to restore causality at the price of introducing non-local effects at the scale $(120 \pi/ G N)^{1/2}$. This was done, for example, in \cite{Donoghue:2014yha} within the context of Friedmann, Lema\^itre, Robertson and Walker cosmology.  The Lee-Wick prescription can also be used to make sense of complex poles \cite{Modesto:2015ozb,Modesto:2016ofr}. The scale of non-locality is thus potentially much larger than $l_P$ if there are many fields in the matter sector, i.e., if $N$ is large.

In \cite{Calmet:2015dpa}, it was shown that the resummed graviton propagator in Eq. (\ref{resprop}) induces non-local effects in scalar field theories at short distances of the order of $(120 \pi/ G N)^{1/2}$. We extend this work to spinor and vector fields and demonstrate that the non-local effects propagate universally in quantum field theory as would be expected from quantum black holes and the thought experiments described previously. We consider a theory with an arbitrary number of scalar fields, spinor and vector fields and calculate their two-by-two scattering gravitational amplitudes using the dressed graviton propagator  (\ref{resprop}). We then extract the leading order (i.e. order $G^2 N$) term of each of these amplitudes and present the results in terms of effective operators.

The stress-energy tensors for the different field species with spins 0, 1/2 and 1 are given as usual by
\begin{eqnarray}
  T_\text{scalar}^{\mu\nu} &= & \partial^\mu\phi ~\partial^\nu \phi - \eta^{\mu\nu} L_\text{scalar} ~,\\
  T_\text{fermion}^{\mu\nu} &=& \cfrac{i}{4} \bar\psi \gamma^\mu \nabla^\nu \psi + \cfrac{i}{4} \bar \psi \gamma^\nu\nabla^\mu \psi - \cfrac{i}{4} \nabla^\mu \bar\psi \gamma^\nu \psi - \cfrac{i}{4} \nabla^\nu \bar\psi \gamma^\mu \psi  - \eta^{\mu\nu}L_\text{fermion} ~, \\
  T_\text{vector}^{\mu\nu} &= & - F^{\mu\sigma} F^\nu_{~~\sigma}  + m^2 A^\mu A^\nu -\eta^{\mu\nu} L_\text{vector} ~,
\end{eqnarray}
where we have used the following free field matter Lagrangians:
\begin{eqnarray}
  L_\text{scalar} & =& \cfrac12 (\partial\phi)^2 - \cfrac12 m^2 \phi^2 ~ , \\
  L_\text{fermion} &=& \cfrac{i}{2} \bar\psi \gamma^\sigma\nabla_\sigma \psi - \cfrac{i}{2} \nabla_\sigma \bar\psi \gamma^\sigma \psi - m \bar\psi\psi ~, \\
  L_\text{vector} &=& -\cfrac14 F^2 +\cfrac12 m^2 A^2 ~.
\end{eqnarray} 

We can now present the complete set of non-local operators at order $NG^2$.  The non-local operators involving scalar fields only are given by
\begin{eqnarray}
  O_\text{scalar,1} &= & \cfrac{N G^2}{30\pi} ~\pd_\mu\phi~\pd_\nu\phi~ \Rlog{\square} ~ \pd^\mu \phi' ~ \pd^\nu\phi' ~, \\
  O_\text{scalar,2} &=& -\cfrac{NG^2}{60\pi} ~\pd_\mu\phi\pd^\mu\phi ~\Rlog{\square} ~\pd_\sigma\phi'\pd^\sigma \phi' ~, \\
  O_\text{scalar,3} &=&\cfrac{NG^2}{30\pi}~ L_\text{scalar} ~\Rlog{\square} ~ \pd_\sigma\phi'\pd^\sigma \phi' ~, \\
  O_\text{scalar,4} &=& \cfrac{NG^2}{30\pi}~ \pd_\mu\phi\pd^\mu\phi ~\Rlog{\square} ~ L'_\text{scalar} ~, \\
  O_\text{scalar,5} &=& -\cfrac{2 NG^2}{15\pi} ~L_\text{scalar} \Rlog{\square} ~L'_\text{scalar} ~.
\end{eqnarray}

The non-local operators involving spinor fields only are given by
\begin{eqnarray}
  O_\text{fermion,1} & =&  \cfrac{NG^2}{60\pi} \left( \cfrac{i}{2} \bar \psi \gamma^\mu \nabla^\nu \psi - \cfrac{i}{2} \nabla^\mu \bar\psi \gamma^\nu \psi \right) \left(\delta_\mu^\alpha \delta_\nu^\beta + \delta_\mu^\beta \delta_\nu^\alpha\right) \Rlog{\square} \left( \cfrac{i}{2} \bar\psi' \gamma^\alpha\nabla^\beta \psi' - \cfrac{i}{2} \nabla^\alpha \bar\psi' \gamma^\beta \psi' \right) ~, \nonumber \\ 
   &&\\
  O_\text{fermion,2}& =&- \cfrac{NG^2}{60\pi} ~ \left(\cfrac{i}{2} \bar\psi\gamma^\sigma\nabla_\sigma \psi - \cfrac{i}{2} \nabla_\sigma\bar\psi\gamma^\sigma\psi \right) \Rlog{\square} \left(\cfrac{i}{2} \bar\psi' \gamma^\rho\nabla_\rho \psi' - \cfrac{i}{2} \nabla_\rho \bar\psi' \gamma^\rho \psi' \right) ~,\\
  O_\text{fermion,3} &=& \cfrac{NG^2}{30\pi} L_\text{fermion} \Rlog{\square} \left( \cfrac{i}{2} \bar\psi'\gamma^\sigma\nabla_\sigma \psi'- \cfrac{i}{2} \nabla_\sigma \bar\psi' \gamma^\sigma \psi'\right) ~,\\
  O_\text{fermion,4} &=& \cfrac{NG^2}{30\pi} \left(\cfrac{i}{2} \bar\psi \gamma^\sigma\nabla_\sigma \psi - \cfrac{i}{2} \nabla_\sigma \bar\psi \gamma^\sigma \psi \right) \Rlog{\square} L'_\text{fermion} ~,\\
  O_\text{fermion,5} & =& - \cfrac{2NG^2}{15\pi} L_\text{fermion} \Rlog{\square} L'_\text{fermion}~.
\end{eqnarray}

The non-local operators involving vector fields only are given by
\begin{eqnarray}
  O_\text{vector,1} &=& \cfrac{NG^2}{30\pi} \left( F^{\mu\sigma}F_{\nu\sigma} - m^2 A^\mu A_\nu \right)\Rlog{\square} \left( F'^{\mu\rho} F'_{\nu\rho} - m'^2 A'_\mu A'^\nu \right)  ~,\\
  O_\text{vector,2} &=& - \cfrac{NG^2}{60\pi} (F^2 - m^2 A^2) \Rlog{\square} (F'^2 - m'^2 A'^2) ~,\\
  O_\text{vector,3} &=& - \cfrac{NG^2}{30\pi} L_\text{vector} \Rlog{\square} (F'^2 - m'^2 A'^2) ~,\\
  O_\text{vector,4} &=& - \cfrac{NG^2}{30\pi} (F^2 - m^2 A^2) \Rlog{\square} L'_\text{vector} ~,\\
  O_\text{vector,5} &=& - \cfrac{2NG^2}{15\pi} L_\text{vector} \Rlog{\square} L'_\text{vector}  ~.
\end{eqnarray}

The non-local operators involving amplitudes with scalar and vector fields only are given by
\begin{eqnarray}
  O_\text{scalar-vector,1} &=&- \cfrac{NG^2}{30\pi} \pd^\mu \phi \pd_\nu\phi ~\Rlog{\square} ~\left( F_{\mu\sigma} F^{\nu\sigma}-m_A^2 A_\mu A^\nu  \right) ~,\\
  O_\text{scalar-vector,2} &=& \cfrac{NG^2}{60\pi} (\pd\phi)^2 \Rlog{\square} (F^2 - m_A^2 A^2) ~, \\
  O_\text{scalar-vector,3} &=&- \cfrac{NG^2}{30\pi} L_\text{scalar} \Rlog{\square} ~(F^2 - m_A^2 A^2 ) ~, \\
  O_\text{scalar-vector,4} &=& - \cfrac{NG^2}{30\pi} (\pd\phi)^2 \Rlog{\square} L_\text{vector} ~, \\
  O_\text{scalar-vector,5} &=& - \cfrac{2NG^2}{15\pi} L_\text{scalar}\Rlog{\square} L_\text{vector} ~.
\end{eqnarray}

The non-local operators involving amplitudes with scalar and spinor fields only are given by
\begin{eqnarray}
  O_\text{scalar-fermion,1} &=& \cfrac{NG^2}{30\pi} \pd_\mu\phi ~\pd_\nu\phi \Rlog{\square} \left(\cfrac{i}{2} \bar\psi\gamma^\mu\nabla^\nu\psi - \cfrac{i}{2} \nabla^\mu\bar\psi\gamma^\nu\psi \right) ~,\\
  O_\text{scalar-fermion,2} &=& -\cfrac{NG^2}{60\pi} ~(\pd\phi)^2 \Rlog{\square} \left(\cfrac{i}{2}\bar\psi\gamma^\sigma\nabla_\sigma\psi - \cfrac{i}{2}\nabla_\sigma\bar\psi\gamma^\sigma\psi \right) ~,\\
  O_\text{scalar-fermion,3} &=& \cfrac{NG^2}{30\pi} ~L_\text{scalar} \Rlog{\square} ~\left(\cfrac{i}{2} \bar\psi \gamma^\sigma\nabla_\sigma \psi - \cfrac{i}{2} \nabla_\sigma\bar\psi\gamma^\sigma \psi\right) ~,\\
  O_\text{scalar-fermion,4} &=& \cfrac{NG^2}{30\pi} ~(\pd\phi)^2 \Rlog{\square}~L_\text{fermion} ~, \\
  O_\text{scalar-fermion,5} &=&- \cfrac{2NG^2}{15\pi} ~ L_\text{scalar} \Rlog{\square} L_\text{fermion} ~.
\end{eqnarray}

The non-local operators involving amplitudes with spinor and vector fields only are given by
\begin{eqnarray}
  O_\text{vector-fermion,1} &=& -\cfrac{NG^2}{30\pi} \left( \cfrac{i}{2} \bar\psi \gamma^\mu \nabla^\nu\psi -\cfrac{i}{2} \nabla^\mu \bar\psi \gamma^\nu \psi \right) \Rlog{\square} \left( F_{\mu\sigma} F_\nu^{~~\sigma} - m_A^2 A_\mu A_\sigma \right) ~,\\
  O_\text{vector-fermion,2} &=& \cfrac{NG^2}{60\pi} \left(\cfrac{i}{2}\bar\psi\gamma^\sigma\nabla_\sigma \psi - \cfrac{i}{2} \nabla_\sigma \bar\psi \gamma^\sigma \psi \right) \Rlog{\square} (F^2 - m_A^2 A^2 ) ~,\\
  O_\text{vector-fermion,3} &=&-\cfrac{NG^2}{30\pi} L_\text{fermion} ~\Rlog{\square} (F^2 -m_A^2 A^2) ~,\\
  O_\text{vector-fermion,4} &=&\cfrac{NG^2}{30\pi} \left(\cfrac{i}{2}\bar\psi\gamma^\sigma\nabla_\sigma \psi \right) \Rlog{\square} L_\text{vector} ~,\\
  O_\text{vector-fermion,5} &=&-\cfrac{2NG^2}{15\pi} L_\text{fermion} \Rlog{\square} L_\text{vector} ~.
\end{eqnarray}

By looking at these effective operators, one can see explicitly that the gravitational non-locality leads to non-local effects in the matter sector as well. This is the case for all matter fields of any spin. The non-locality is manifest due to the presence of the $\log(\square)$ term in all of these effective operators. The universality of the non-locality in the matter sector is precisely what one expects in the context of a minimal length. The underlying argument in all minimal length demonstrations is the following. When length scales shorter than the minimal length are probed, one ends up concentrating so much energy within that region of space-time that a Planckian black hole will eventually form in that region of space-time. This is precisely what we are finding when interpreting the complex pole as a black hole precursor which is an extended object of size $(120 \pi/ G N)^{1/2}$. Its extension in space corresponds to the minimal length that can be probed. We conclude that space-time is smeared on distances shorter than the dynamical Planck scale given by $M_\star=M_P\sqrt{120 \pi/N}$ which corresponds to the energy of the complex pole.

This non-locality prevents an observer from testing distances shorter than the corresponding length scale. It also implies that singularities cannot be probed experimentally as space-time is smeared. One may argue that the notion of space-time looses its meaning on distances smaller than $1/M_\star$. This interpretation fits well with the observations made recently in \cite{Calmet:2017qqa}.

It is interesting to point out that the non-local effects in the four-fermion interactions can be probed at the Large Hadron Collider.    The ATLAS collaboration has searched for four-fermion contact interactions at $\sqrt{s}=8$ TeV and obtained lower limits on the scale on the lepton-lepton-quark-quark contact interaction $\Lambda$ between 15.4 TeV and 26.3 TeV \cite{Aad:2014wca}. The most restrictive bound on $\Lambda$ is obtained by combining the dielectron and dimuon channels. We have contributions to these process coming from $O_\text{fermion,1}$ and  $O_\text{fermion,2}$. We first note that the renormalization scale $\mu$ should scale with $N$ as well, we take $\mu^2=120\pi/( N G)$. Since we are looking at conservative order of magnitudes, we will identify the scale generated by the derivatives in the four-fermion operators with the center of mass energy of the proton-proton collision. We are thus dealing effectively with operators of the type $\bar q q \bar l l $ which are suppressed by a factor $2 N G^2 /(60\pi^2) s \log(s N G/(120 \pi))$. This translates into a conservative bound $N<5\times10^{61}$ on the number of light fields in a hidden sector. This implies that the scale $M_\star$, which parametrizes the  non-locality of space-time, is larger than $3\times 10^{-11}$ GeV. This bound is tighter than that obtained from gravitational waves and from E\"otv\"os type pendulum experiments \cite{Calmet:2016sba} by two orders of magnitude.

Note that our bound on the scale of space-time non-locality ($M_P\sqrt{120\pi/N}$) is much weaker than those on the Planck mass ($M_P$) obtained using the standard geometrical cross section (i.e. $\sigma=\pi R_S^2$ where $R_S$ is the Schwarzschild radius)  for quantum black holes \cite{bh1,bh2,bh3,Calmet:2008dg,Calmet:2011ta,Calmet:2012fv,Alberghi:2013hca,Belyaev:2014ljc,sa1,sa2,sa3,sa4,sa5,sa6,sa7,sv1,Sirunyan:2017anm}. Collider bounds on a new object with a geometric cross section are typically of the order of 9 TeV \cite{Sirunyan:2017anm}. This is not surprising as we are indeed studying different higher order effective operators. So far, we have not found, within the effective theory approach, higher dimensional operators corresponding to the geometrical cross section which has been extensively studied. The intermediate states in the propagator of the graviton, which we have studied, couple with the usual Planck mass to the particles of the standard model while in the more extensively studied models, quantum black holes are assumed to couple much stronger (i.e. with $M_P \sim$ TeV) to the particles of the standard model. Since at this stage we have not identified a mechanism which lowers the value of the Planck mass (we are using dimensional regularization in contrast to \cite{Calmet:2008tn} where a dimensionful cutoff had been used), there is no strong gravitational effect in the TeV region to be expected.
 
It is worth mentioning that weakly nonlocal theories, such as the effective field theory for general relativity considered in this paper, can be also the starting point to construct an ultraviolet completion of Einstein's gravity, which turns out to be unitary at perturbative level and finite at quantum level \cite{Modesto:2011kw,Modesto:2014lga,Modesto:2015lna}.

In this paper, we have shown that the non-locality recently identified in quantum gravity propagates to the matter sector of the theory. We have described these non-local effects using the tools of effective field theory. We have derived the complete set of effective operators at order $N G^2$ for theories involving scalar, spinor, and vector fields. We then have used recent data from the Large Hadron Collider to set a bound on the scale of space-time non-locality and found $M_\star>3\times 10^{-11}$ GeV.

{\it Acknowledgments:}
The work of XC is supported in part  by the Science and Technology Facilities Council (grant number  ST/P000819/1). XC is very grateful to PRISMA and the MITP for their generous hospitality during the academic year 2017/2018. The work of BL and SA is supported in part by the Russian Foundation for Basic Research via grant number 16-02-00682. 


\bigskip{}

\end{document}